\documentclass[
prc,amsmath,amssymb,floatfix,nofootinbib]{revtex4}

\usepackage{graphicx}
\usepackage{psfrag}

\def\ve{\varepsilon}
\def\Re{\mathop{\rm Re}\nolimits}
\def\Im{\mathop{\rm Im}\nolimits}
\def\CG{{\cal G}}
\def\CM{{\cal M}}

\begin{document}
\title{Two-photon exchange in dispersion approach}
\author{Dmitry~Borisyuk} 
\author{Alexander~Kobushkin} 
\affiliation{Bogolyubov Institute for Theoretical Physics\\
Metrologicheskaya street 14-B, 03680, Kiev, Ukraine}
\date{\today}
%
\begin{abstract}
We calculate two-photon exchange amplitude for the elastic electron-proton
scattering in the framework of dispersion relations.
The imaginary part of the amplitude is determined by unitarity.
Since in the unitarity relation intermediate states are on shell,
off-shell form factors are not needed for the calculation.
The real part is then evaluated using analytical properties of the amplitude.
The expression for the elastic contribution to the amplitude,
obtained in our approach, differs from the results of traditional calculations
with on-shell form factors.
Nevertheless, numerically the difference is minor
for $Q^2$ up to 6 GeV$^2$.
\end{abstract}
\maketitle
\section{Introduction}
The precision level of present-day electron-proton scattering experiments
makes it necessary to take into account effects beyond Born approximation,
such as two-photon exchange (TPE). TPE can be seen in various observables
in wide kinematical range; in particular it influences proton radius
measurements \cite{radius},
generates non-zero transverse beam spin asymmetry \cite{BNSA},
and, the most important, TPE corrections play crucial role in reconciliation
of different measurements of proton form factors (FFs) at high $Q^2$ \cite{Arrington}.
Clearly, such corrections are also required for analysis of data from
upcoming measurements at higher $Q^2$ \cite{expts}.

\begin{figure}
 \psfrag{k}{$k$}\psfrag{k'}{$k'$}\psfrag{k''}{$k''$}\psfrag{p}{$p$}\psfrag{p'}{$p'$}
\includegraphics[width=0.25\textwidth]{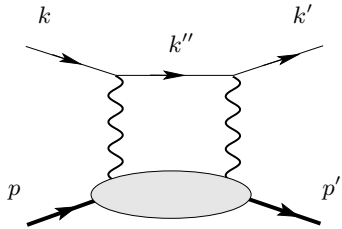}
 \caption{TPE diagram}\label{tpe}
\end{figure}

The TPE diagram (Fig.~\ref{tpe}) for elastic $ep$ scattering differs from similar
diagram in QED in two ways. First, the proton is not a point-like object,
thus there are some non-trivial FFs at $\gamma p$ vertices.
Second, the interaction of the proton with virtual photon may lead to
excitation of inelastic intermediate states, such as $\pi p$,
$\Delta$ resonance, and so on.

At present, the calculations exist for elastic intermediate state
\cite{ourbox,BMTbox}, and for a number of resonances \cite{BMTres}%
\footnote{%
There are also partonic model calculations, appropriate at large
$Q^2$ and energy \cite{partonic}.
This approach will not be considered further.}%
.
Though the evaluation of loop integral in these papers was almost perfect,
the weak point of all such calculations is the starting expression
for the TPE diagram (here we consider the elastic contribution,
but the contribution of resonances may be studied similarly).
This expression results from the contraction of ``leptonic'' and ``hadronic''
parts
\begin{equation} \label{o}
 i\CM^{(\rm naive)} = \int \frac{(4\pi\alpha)^2}{q_1^2 q_2^2}
 L^{\mu\nu} H_{\mu\nu} \frac{d^4 k''}{(2\pi)^4}
\end{equation}
where $q_1 = k-k''$, $q_2 = k''-k'$,
the leptonic part $L_{\mu\nu}$ comes from QED
\begin{equation} \label{oo}
 L_{\mu\nu} = \bar u' \gamma_\mu \,\frac{\hat k'' + m}{k''^2 - m^2}\, \gamma_\nu u
\end{equation}
but the hadronic part $H_{\mu\nu}$ is only $guessed$ to be
\begin{equation} \label{ooo}
H_{\mu\nu} = \bar U' \Gamma_\mu(q_2) \,\frac{\hat p''+M}{p''^2-M^2}\,
   \Gamma_\nu(q_1) U 
   + \bar U' \Gamma_\nu(q_1) \,\frac{\hat{\tilde p}''+M}{\tilde p''^2-M^2}\,
   \Gamma_\mu(q_2) U
\end{equation}
%
where $p''=p+q_1$, $\tilde p''=p+q_2$ and
$\Gamma_\mu(q)$ is the amplitude of proton interaction 
with electromagnetic field, written in the form
\begin{equation}
 \Gamma_\mu(q) = \gamma_\mu F_1(q^2) -
   \frac{1}{4M} F_2(q^2) [\gamma_\mu,\hat q]
\end{equation}
The above-described form of the TPE amplitude was used by many authors
from Bodwin and Yennie in 1988 \cite{Yennie}
to the latest papers \cite{ourbox,BMTbox}.
The justification for such choice of $H_{\mu\nu}$ is the following:
first, it gives the expected result if the intermediate proton is on-shell,
or more precisely, it has correct residues at $p''^2 = M^2$ and
$\tilde p''^2 = M^2$, and second, this expression is gauge-invariant, i.e.
\begin{equation}
 q_{1\nu} H_{\mu\nu} = q_{2\mu} H_{\mu\nu} = 0
\end{equation}

However, (\ref{ooo}) is not the only expression with such properties. One can,
for instance, add to $F_2$ an arbitrary function which vanishes at
$p''^2=M^2$, like
\begin{equation}
 F_2(q^2) \to F_2(q^2) + (p''^2-M^2)f(q^2)
\end{equation}
We should emphasize that, if we are dealing with the elastic contribution only,
than the choice of $H_{\mu\nu}$ is somewhat a matter of convention,
since any change of the elastic contribution may be compensated
by appropriate redefinition of the inelastic one.
Nevertheless, it is desirable to have clear, unambiguous,
and easy-for-calculation definition for contribution
of each intermediate state.

The modification of $\gamma p$ vertex at $p''^2 \ne M^2$,
which of course can be more general than the example shown above,
is usually referred to as introduction of proton off-shell FFs.
The uncertainty of these FFs is believed to be the main source of theoretical
uncertainty in TPE amplitudes \cite{Arrington}.
On the other hand, such FFs are not directly measurable,
just because off-shell proton cannot be a final state.
Hence to take into account off-shell behaviour, one cannot rely
on experimental data but instead must use some nucleon model.
This is undesirable, since the result will be model-dependent.

In the current paper we propose a consistent approach to calculation of TPE,
in which the use of ``off-shell'' FFs is avoided.
The approach is based on the dispersion relations.
At first, the absorptive part of the amplitude is calculated using unitarity.
Thus only ``on-shell'' FFs are needed to evaluate it.
Then the whole amplitude is reconstructed by dispersion relations.
Since this operation is linear, contributions from different intermediate states
may be treated separately.
\section{The amplitudes}
We follow the notation of Refs.\cite{ourbox,ourLow,ourPheno}.
In particular, we define $P=(p+p')/2$, $K=(k+k')/2$ and
$t = q^2$, $\nu = s-u = 4PK$, where $s$, $t$ and $u$ are Mandelstam variables.
The electron and proton masses are $m$ and $M$, respectively.
In the present section (but not in the whole paper)
the electron mass is neglected.

The general-case elastic $ep$ scattering amplitudes is conveniently
written as \cite{GV}
\begin{equation}\label{calM}
  {\cal M} = \frac{4\pi\alpha}{q^2} \bar u'\gamma_\mu u \cdot
 \bar U' \left(\gamma^\mu \tilde F_1
 - \frac{1}{4M} [\gamma^\mu,\hat q] \tilde F_2
 + \frac{P^\mu}{M^2} \hat K \tilde F_3
 \right) U
\end{equation}
In Ref.~\cite{ourLow} the following set of amplitudes was introduced
\begin{equation}
 \begin{array}{ccl}
  \CG_E & = & \tilde F_1 - \tau \tilde F_2 + \nu \tilde F_3/4M^2\\
  \CG_M & = & \tilde F_1 +\tilde F_2 + \ve \nu \tilde F_3/4M^2\\
  \CG_3 & = & \nu \tilde F_3/4M^2
 \end{array}
\end{equation}
which ``diagonalizes'' the cross-section
\begin{equation} \label{...}
 d\sigma = \frac{2\pi\alpha^2 dt}{E^2 t} \frac{1}{1-\ve}
 \left( \ve |\CG_E|^2 + \tau |\CG_M|^2 +
   \tau \ve^2 \frac{1-\ve}{1+\ve} |\CG_3|^2 \right)
\end{equation}
In the above equations, $E$ is initial electron lab. energy,
$\tau = -t/4M^2$ and $\ve = [\nu^2+t(4M^2-t)]/[\nu^2-t(4M^2-t)]$.
Since the amplitude $\CG_3$ vanishes in Born approximation
and hence is $O(\alpha)$, the last term in (\ref{...}) is negligibly small
and we have
\begin{equation} \label{...x}
 d\sigma \approx d\sigma_0 \left( \ve |\CG_E|^2 + \tau |\CG_M|^2 \right)
\end{equation}
similarly to Rosenbluth formula, except that $\CG_E$ and $\CG_M$ are
$\ve$-dependent.

However to make use of the dispersion relations, we need amplitudes,
free from kinematical $u$ and $s$ singularities and zeros.
Such amplitudes are easily constructed by consideration
of annihilation channel. The helicity amplitudes of the process
$e^- e^+ \to p \tilde p$ are
\begin{eqnarray}
 &&T_{++} =  4\pi\alpha\cdot 2i\cos^2\theta/2 \left(
   \sqrt{\tau(1+\tau)} \tilde F_3 + \tilde F_m + \nu\tilde F_3/4M^2
   \right)\\
 &&T_{--} =  4\pi\alpha\cdot 2i\sin^2\theta/2 \left(
   \sqrt{\tau(1+\tau)} \tilde F_3 - \tilde F_m - \nu\tilde F_3/4M^2
   \right)\\
 &&T_{+-} = T_{-+} =  4\pi\alpha\cdot \frac{2M}{\sqrt{t}} \sin\theta \left(
  \tilde F_e + \nu\tilde F_3/4M^2 \right)
\end{eqnarray}
where $\tilde F_e = \tilde F_1 - \tau \tilde F_2$,
$\tilde F_m = \tilde F_1 + \tilde F_2$, and
$\theta$ is $t$-channel scattering angle,
\begin{equation}
 \cos\theta = -\nu/\sqrt{-t(4M^2-t)}
\end{equation}
The subscripts of the quantity $T_{\lambda\tilde\lambda}$ indicate
the signs of proton and antiproton helicities, respectively,
while the electron and positron helicities are $+1/2$ and $-1/2$.
Computing the scattering channel cross-section
\begin{equation}
 \frac{d\sigma}{dt} = \frac{1}{64\pi M^2 E^2} \cdot \frac{1}{2}
  \left( |T_{++}|^2 + |T_{--}|^2 + 2|T_{+-}|^2 \right)
\end{equation}
we return to the formula (\ref{...}).
Each of the $T_{\lambda\tilde\lambda}$ contains a kinematical factor of
$\sin^{|\lambda+\tilde\lambda-1|}\frac{\theta}{2}
 \cos^{|\lambda+\tilde\lambda+1|}\frac{\theta}{2}$ (see e.g. Ref.~\cite{Wang}).
The amplitudes free from kinematical singularities are obtained
after removing these factors, i.e.
\[
  \sqrt{\tau(1+\tau)} \tilde F_3 \pm
  \left(\tilde F_m + \nu\tilde F_3/4M^2 \right)
  \qquad {\rm and} \qquad
  \tilde F_e + \nu\tilde F_3/4M^2 
\]
or equivalently
\begin{equation}\label{Gn}
 G_1 \equiv \CG_E = \tilde F_e + \nu\tilde F_3/4M^2,\qquad
 G_2 = \tilde F_m + \nu \tilde F_3/4M^2,\qquad
 G_3 \equiv \tilde F_3
\end{equation}
The amplitudes $G_n$ satisfy fixed-$t$ dispersion relations
\begin{equation} \label{d}
 \pi G_n(\nu) = \int\limits_{\nu_{th}}^\infty \frac{\Im G_n(\nu'+i0)}{\nu'-\nu}d \nu'
 - \int\limits^{-\nu_{th}}_{-\infty} \frac{\Im G_n(\nu'-i0)}{\nu'-\nu} d\nu'
\end{equation}
and consequently, vanish at $\nu\to\infty$.
Under crossing $\nu \to -\nu$ two first amplitudes are odd and the last
is even:
\begin{equation} \label{dd}
 G_{1,2}(-\nu) = - G_{1,2}(\nu),\qquad G_3(-\nu) = G_3(\nu).
\end{equation}
\section{Calculation procedure}
\subsection{Imaginary part}
The imaginary part of the scattering amplitude can be calculated
via unitarity condition
\begin{equation}
 T_{if}^* - T_{fi} = \sum_n T_{fn} T_{in}^*
\end{equation}
or graphically
\begin{equation} \label{u}
 \psfrag{h}{$\bigr\}h$}\psfrag{2Im}{$2\Im$}
\psfrag{=}{ $\displaystyle = \ \int\frac{d^3 \vec k''}{2k''_0} \sum\limits_h$}
\psfrag{k}{$k$}\psfrag{k'}{$k'$}\psfrag{k''}{$k''$}\psfrag{p}{$p$}\psfrag{p'}{$p'$}\psfrag{x}{$\times$}
 \includegraphics[width=0.5\textwidth]{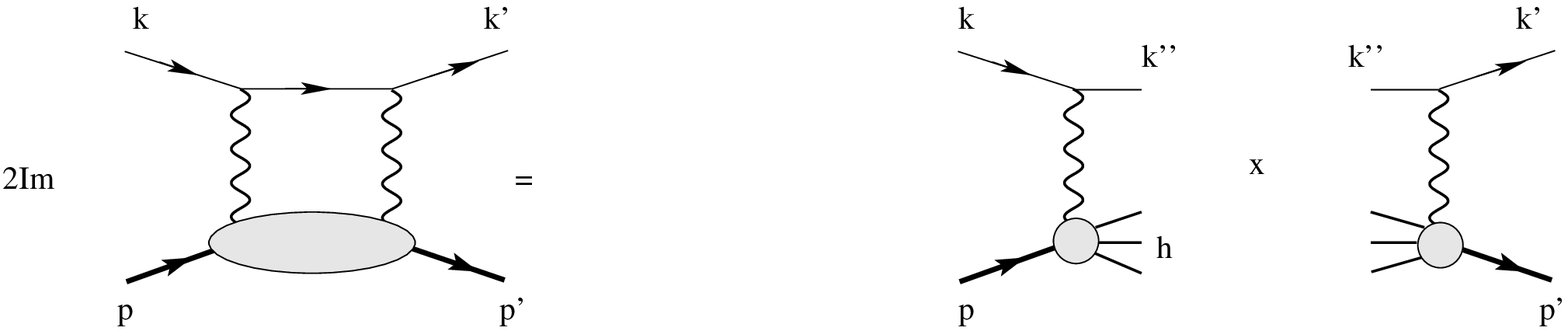}
\end{equation}
where we have replaced $T$-matrix elements in the r.h.s. by their Born
(one-photon exchange) approximations. Thus obtained is exactly the absorptive
part of the TPE amplitude.

Eq.(\ref{u}) allows for natural and unambiguous classification of different
contributions to $\Im G_n$, according to intermediate hadronic states $h$.
The term with $h={\rm proton}$ will be called elastic contribution,
the term with $h=\Delta(1232)$ will be the $\Delta$ resonance
contribution and so on.
Since the intermediate states appearing in the unitarity condition are
real (``on-shell'') particles, it is sufficient to know
on-shell transition amplitudes of these states to calculate $\Im G_n$.
Thus in particular the knowledge of proton ``off-shell'' FFs is not needed.

The reconstruction of the $\Re G_n$ from $\Im G_n$ by dispersion integral
is linear operation, therefore we may introduce a natural definition
of elastic contribution to the whole amplitude as the quantity
yielded by dispersion relation applied to the elastic part of $\Im G_n$,
and similarly for other contributions.
\subsection{Reconstruction of the real part}
From now on we consider the elastic contribution only.
Such contribution to the imaginary part of invariant amplitudes $G_n$
can be written in the form
\begin{equation} \label{*}
 \Im G_n^{(\rm el)} = - \frac{\alpha}{2\pi} \sum_{i,j=1}^2 \int \bar F_i(t_1) \bar F_j(t_2)
  A_{n,ij}(\nu,t_1,t_2)
  \theta(k''_0)\delta(k''^2-m^2) \theta(p''_0)\delta(p''^2-M^2) d^4 k''
\end{equation}
where $\bar F_i(t) = F_i(t)/(t-\lambda^2)$, $A_{n,ij}$ is a polynomial in
$t_1$, $t_2$ and a rational function
of $\nu$ (it may have poles in $\nu$; the explicit expression for $A_{n,ij}$
is given in Appendix~\ref{AppA}). The $\theta$- and $\delta$- functions ensure that
intermediate particles are on-shell. The straightforward way
of further calculation is to insert $\Im G_n^{(\rm el)}$
into the dispersion integral (\ref{d}) and evaluate it.
However this is not an easy task.
For example, Eq.(\ref{*}) for imaginary part is valid only in physical region
$\nu \ge \sqrt{-t(4M^2-t)}$, but the dispersion integral involves all $\nu$
values above the threshold $\nu = \nu_{th}$ (corresponding to $s=(M+m)^2$),
thus before it can be evaluated
we must first find an analytical continuation of (\ref{*})
into the unphysical region. Though such analytical continuation is unique,
it is hard to write it down in a compact form.
So we will use an easier roundabout way.

The amplitude $G_n$ is an analytical function of $\nu$ with two branch cut
discontinuities along the real axis:
from $-\infty$ to $-\nu_{\rm th}$ and from $\nu_{\rm th}$ to $+\infty$.
As implied by Eq.(\ref{d}), it can be written as a sum of two parts,
direct and crossed box amplitudes
\begin{equation}
 G_n(\nu) = G_{n,\rm box}(\nu) + G_{n,\rm xbox}(\nu)
\end{equation}
with each of them having only one discontinuity,
box from $\nu_{\rm th}$ to $+\infty$
and crossed box from $-\infty$ to $-\nu_{\rm th}$.
Direct and crossed box amplitudes are related by
\begin{equation}
 G_{n,\rm box} = \pm G_{n,\rm xbox}(-\nu)
\end{equation}
where $\pm$ is chosen according to (\ref{dd}).
Thus to reconstruct $G_n$ it is sufficient to find $G_{n,\rm box}$.

To do this, we note that if we find any function with the following properties:\\
1) it has no singularities except the branching point at $s=(M+m)^2$,\\
2) its branch cut discontinuity is $2i\Im G_n^{(\rm el)}$,
 with $\Im G_n^{(\rm el)}$ given by Eq.(\ref{*}),\\
3) it vanishes as $s \to \infty$,\\
then such function necessarily coincides with the sought amplitude
(otherwise their difference would be non-trivial bounded whole function,
which is impossible).

The analytical structure of FFs is such that
\begin{equation} \label{**}
 \bar F_i(t) = \frac{1}{\pi} \int\limits_{\lambda^2}^\infty
   \frac{\Im \bar F_i(t')}{t'-t} dt'
\end{equation}
in other words, the FFs in the Eq.(\ref{*}) are some linear combinations
of a single poles $\frac{1}{t-a}$ with $a>0$.
Using the decomposition (\ref{**}), we may obtain
\begin{equation} \label{x}
 \sum_{i,j=1}^2 \bar F_i(t_1) \bar F_j(t_2) A_{ij}(\nu,t_1,t_2) =
 \int\limits_{\lambda^2}^\infty da \int\limits_{\lambda^2}^\infty db \,
 \frac{c(\nu,a,b)}{(t_1 - a)(t_2 - b)}
\end{equation}
and rewrite Eq.(\ref{*}) as
\begin{equation} \label{***}
 \Im G_n^{(\rm el)} = \int\limits_{\lambda^2}^\infty da \int\limits_{\lambda^2}^\infty db \,
 c(\nu,a,b) \int \frac{1}{(t_1-a)(t_2-b)}
 \theta(k''_0)\delta(k''^2-m^2) \theta(p''_0)\delta(p''^2-M^2) d^4 k''
\end{equation}
Consider the function
\begin{equation} \label{i}
 I_4(s,t;a,b) = \int \frac{i\, d^4 k''}{(t_1-a)(t_2-b)(k''^2-m^2)(p''^2-M^2)}
\end{equation}
It is well-known that this is an analytic function of $s$ everywhere
except the branch cut from $s=(M+m)^2$ to $+\infty$.
Its discontinuity across the cut is
\begin{equation}
 \Delta I_4 = 2 i \Im I_4 = \int \frac{-4i \pi^2}{(t_1-a)(t_2-b)}
  \theta(k''_0)\delta(k''^2-m^2) \theta(p''_0)\delta(p''^2-M^2) d^4 k''
\end{equation}
which is exactly the innermost integral in (\ref{***}).
Thus if the coefficients $c(\nu,a,b)$ were independent of $\nu$,
the whole TPE amplitude would be obtained by substitution
\begin{equation} \label{subs}
 \theta(k''_0)\delta(k''^2-m^2) \theta(p''_0)\delta(p''^2-M^2) 
 \to \frac{1}{2i\pi^2}\frac{1}{(k''^2-m^2)(p''^2-M^2)}
\end{equation}
under the integral, yielding
\begin{equation} \label{+}
 \tilde G_n^{(\rm el)}(\nu) = \frac{i\alpha}{4\pi^3}
  \sum_{i,j=1}^2 \int \bar F_i(t_1) \bar F_j(t_2) A_{n,ij}(\nu,t_1,t_2)
    \frac{d^4 k''}{(k''^2-m^2)(p''^2-M^2)}
\end{equation}

But actually quantities $A_{n,ij}(\nu,t_1,t_2)$ and thus $c(\nu,t_1,t_2)$
have poles at the boundary of the physical region
$\nu = \pm\nu_0 = \pm\sqrt{-t(4M^2-t)}$ (see explicit expressions
in Appendix~\ref{AppA}). Because of this function $\tilde G_n^{(\rm el)}$,
constructed by Eq.(\ref{+}), satisfy conditions 2) and 3) but don't satisfy 1)
since it has unphysical poles at $\nu = \pm\nu_0$.

To remove these poles we may simply subtract the principal part of
$\tilde G_n$ Laurent series expansion about $\nu=\pm\nu_0$
\begin{equation} \label{++}
 G_{n,\rm box}^{(\rm el)}(\nu) = \tilde G_n^{(\rm el)}(\nu)
 - \sum_{r=0}^{N-1} \frac{1}{r!}\frac{g_{r+}}{(\nu-\nu_0)^{N-r}}
 - \sum_{r=0}^{N-1} \frac{1}{r!}\frac{g_{r-}}{(\nu+\nu_0)^{N-r}}
\end{equation}
where $N$ is degree of the pole (actually 1 or 2) and
\begin{equation}
 g_{r\pm} = \left. \frac{\partial^r}{\partial\nu^r}(\nu\mp\nu_0)^N
 \tilde G_n^{(\rm el)}(\nu) \right|_{\nu=\pm\nu_0}
\end{equation}
Since the subtracted function is meromorphic (has no branching points)
and vanish at $\nu=\infty$, the properties 2) and 3) hold true
and in addition, the obtained function $G_n^{(\rm el)}(\nu)$ is regular
at $\nu=\pm\nu_0$. So the requirement 1) is also satisfied.
Therefore $G_{n,\rm box}^{(\rm el)}(\nu)$ is the sought amplitude.


In summary, the evaluation of the TPE amplitude proceeds as follows:\\
1) construct the expression for the imaginary part in the form (\ref{*}).\\
2) obtain the quantity $\tilde G_n$, Eq.(\ref{+}),
  by substitution according to Eq.(\ref{subs}).\\
3) subtract unphysical poles at $\nu=\pm\nu_0$, Eq.(\ref{++}).\\
4) perform (anti)symmetrization with respect to $\nu$, i.e. add crossed box amplitude.

Due to decomposition (\ref{x}) the quantity $\tilde G_n$ can be written
as a linear combination of functions $I_4(s,t;a,b)$ with different $a$ and $b$.
This is especially useful if FFs are parameterized as a discrete sum
of a single poles (such an approach was used in Ref.~\cite{BMTbox}).
To perform the subtraction of unphysical poles one needs to know
the value of function $I_4$ and its derivative at $\nu =\pm\nu_0$.
They can be expressed via integrals similar to (\ref{i}) with
$k''-m^2$ or $p''^2-M^2$ or both dropped (such integrals were denoted
$I_1$, $I_2$, $I_3$ in Ref.~\cite{ourbox}). Some useful relations between them
are given in Appendix~\ref{AppB}. With these relations, one may compare
the expression for elastic part of TPE amplitude, obtained
in the dispersion approach, with the ``naive'' result (\ref{o}-\ref{ooo}).
\section{Results and conclusions}
After performing the above-described procedure, we have obtained
the following results for the elastic contributions
to the invariant amplitudes $G_n$. The expressions for $G_1$ and $G_2$
remain the same as in the ``naive'' approach, Eqs.(\ref{o}-\ref{ooo})
\begin{equation}
 G_1 = G_1^{(\rm naive)},\qquad G_2 = G_2^{(\rm naive)}
\end{equation}
The expression for $G_3$ is different:
\begin{equation} \label{tri}
 G_3 = G_3^{(\rm naive)} + \Delta G_3(t)
\end{equation}
where
\begin{equation}
 \Delta G_3(t) = \frac{i\alpha}{4\pi^3 t} \int \frac{F_2(t_1) F_2(t_2)}{t_1\, t_2}
 \left(t_1+t_2+3t-\frac{2 t_1 t_2}{k''^2-m^2} \right) d^4 k''
\end{equation}
The whole scattering amplitude may be written as
\begin{equation}
 \CM = \CM^{(\rm naive)} + \frac{4\pi\alpha}{q^2 M^2}\,
 \bar u'\gamma^\mu u \, \bar U' (P_\mu \hat K - PK \gamma_\mu) U \cdot \Delta G_3(t)
\end{equation}
Since the quantities that contribute to the cross-section
up to the order $O(\alpha)$ are
\begin{equation}
 \CG_E = G_1 \qquad {\rm and} \qquad \CG_M = G_2 - \frac{\nu}{4M^2}(1-\ve) G_3
\end{equation}
(see Eq.(\ref{...x})), with new expression (\ref{tri}) for $G_3$ TPE corrections
to the cross-section will differ from those in ``naive'' approach, since
\begin{equation}
 \CG_M = \CG_M^{(\rm naive)}
 - \sqrt{\tau(1+\tau)}\sqrt{1-\ve^2} \Delta G_3(t)
\end{equation}
Moreover, the affected amplitude, $\CG_M$, is exactly the quantity
which is responsible for the discrepancy between Rosenbluth
and polarization transfer methods in the measurements of proton FFs \cite{ourPheno}.

\begin{figure}[t]
\parbox[t]{0.48\textwidth}{\includegraphics[width=0.48\textwidth]{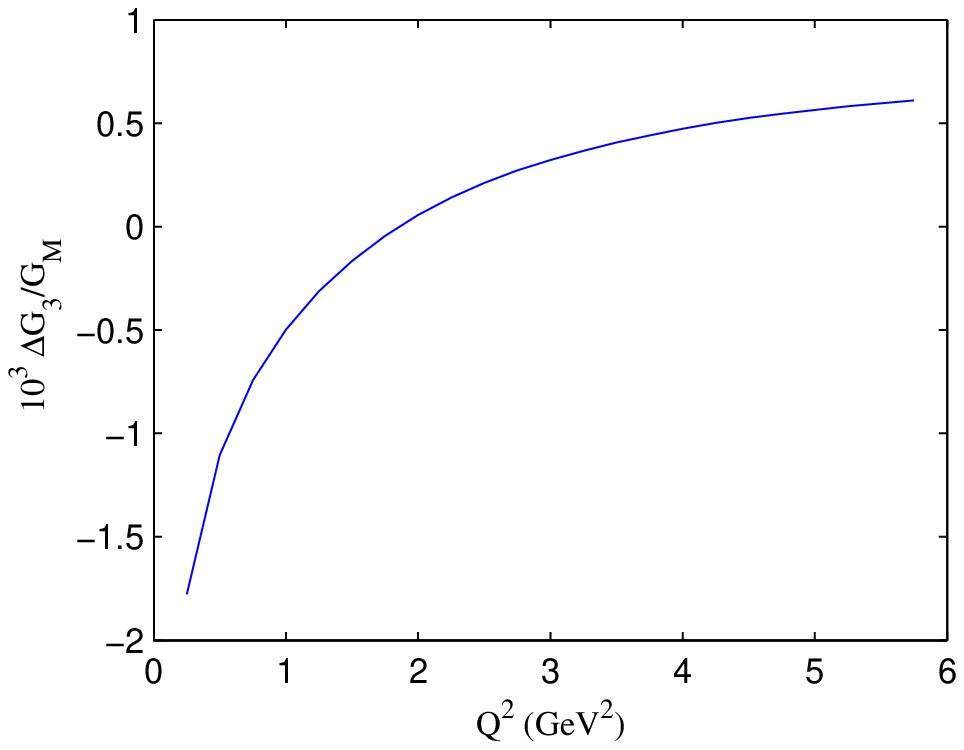}
 \caption{The amplitude change $\Delta G_3$.}\label{deltaG}}
\parbox[t]{0.48\textwidth}
{ \includegraphics[width=0.48\textwidth]{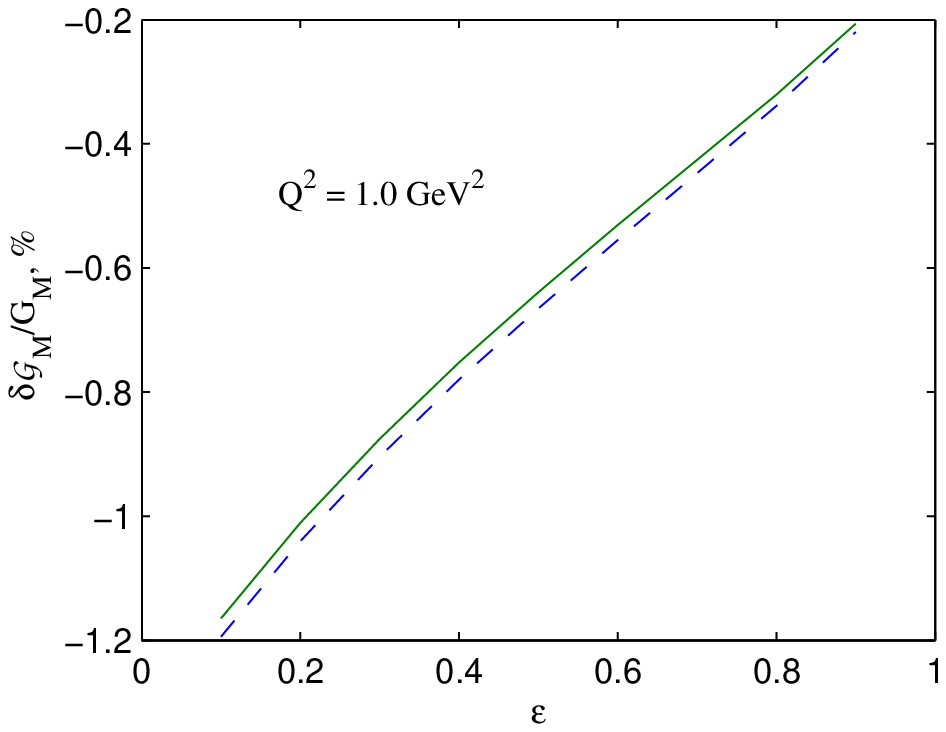}
 \caption{The TPE amplitude $\delta\CG_M$ obtained in old (dashed line) and
 new (solid line) approach.}\label{deltaGM}}
\end{figure}

The numerical calculation, however, shows that the addition to $\CG_M$
is very small (Figs.~\ref{deltaG} and \ref{deltaGM}). Therefore
most of the results obtained starting from ``naive'' expression
for the amplitude will remain unchanged.
In particular, we checked what the low-$Q^2$ behaviour
is the same as reported in Ref.\cite{ourLow},
since the addition to $\CG_M$ vanishes at $Q^2\to 0$.
Nevertheless, the proton off-shell form factors problem is overcome:
they are not needed to calculate TPE amplitudes in our approach.
\appendix
\section{}\label{AppA}
\def\mtx#1#2#3#4{\left(\begin{array}{cc}#1&#2\\#3&#4\end{array}\right)}
The coefficients $A_{n,ij}$ may be computed in the following way.
First, we write down the standard expression for the absorptive part
of the amplitude (elastic contribution)
\begin{eqnarray}
  \Im\CM = \frac{1}{8\pi^2} \int \frac{(4\pi\alpha)^2}{q_1^2 q_2^2}
    \bar u' \gamma_\mu (\hat k'' + m) \gamma_\nu u \cdot
    \bar U' \Gamma_\mu(q_2) (\hat p''+M) \Gamma_\nu(q_1) U \times \nonumber \\
  \times \theta(k''_0)\delta(k''^2-m^2) \theta(p''_0)\delta(p''^2-M^2) d^4k''
\end{eqnarray}
Then we decompose it into scalar invariant amplitudes according to Eq.(\ref{calM})
(the equation for $\Im\CM$ contains $\Im \tilde F_n$ instead of $\tilde F_n$
in the r.h.s.).
Since $\Im \tilde F_n$ are scalars, they will depend on scalar combinations of
$p$, $k$, $p'$, $k'$, $p''$ and $k''$. But due to "onshellness"
of the intermediate particles we have $p''^2=M^2$ and $k''^2=m^2$,
and other scalar products can be expressed via $q_1^2$, $q_2^2$, $\nu$ and $t$.
The vertex functions $\Gamma_\mu$ and $\Gamma_\nu$ contain FFs,
so the resulting expression will be quadratic in FFs.
The $\Im G_n$ are obtained as linear combinations of $\Im \tilde F_n$,
Eq.(\ref{Gn}).

Below $A_{n,ij}$ are written in a matrix notation,
$A_n = \mtx{A_{n,11}}{A_{n,12}}{A_{n,21}}{A_{n,22}}$.
In these formulas $t_p=t_1+t_2-t$, $t_m=t_1-t_2$ and
$\nu_0^2 = -t(4M^2-t)$.
\begin{eqnarray}
 &&A_1 = t ( \nu  - t )  
    \left\{ \frac{1}{2} + 
      t_p \frac{4 M^2 + 2 \nu  - t + t_p}{4 (\nu^2 - \nu_0^2)}
    \right\}
\mtx2000 +
   \frac{t t_m (\nu - t) }{16 M^2}
   \left\{ 2 + t_p \frac{4 M^2 + \nu  - t}{\nu^2 - \nu_0^2}
\right\} \mtx0{-1}10 +
\nonumber \\[1mm]  &&
+ \frac{t (\nu - t) }{16 M^2} 
    \left\{ 2 t + t_p + 
      t_p \frac{( 2 t + t_p ) ( 4 M^2 + \nu  - t ) + 2 M^2 t_p}
      {\nu^2 - \nu_0^2}
\right\}
\mtx0110 +
\nonumber \\[1mm]  &&
+ \left\{
   \frac{t t_p (\nu - t) ( 2 M^2 t_p - \nu (\nu - t) ) }
       {16 M^2 ( \nu^2 - \nu_0^2) } - 
       \frac{t t_1 t_2}{4 M^2}
\right\}
\mtx0112
\\[5mm]
 &&A_2 = (\nu - t)  
     \left\{ \frac{t}{2} + 
       t_p \frac{t ( 4 M^2 + \nu  - t )  - t_p ( 2 M^2 - t ) }
           {2 ( \nu^2 - \nu_0^2 ) } + 
       t t_p^2 \frac{( 4 M^2 - t ) ( 2 M^2 + \nu  - t ) }
           {(\nu^2 - \nu_0^2 )^2}
\right\} \mtx2110 -
\nonumber \\[1mm]  &&
   - \frac{t (\nu - t) t_p}{16 M^2}
   \left\{ 1 + 
       \frac{( 4 M^2 - t + t_p)(\nu - t) + 2 M^2 t_p}
       {\nu^2 - \nu_0^2} +
       2 ( 4 M^2 - t )(\nu - t) t_p
        \frac{ 2 M^2 + \nu - t }{(\nu^2 - \nu_0^2)^2}
\right\} \mtx0002 +
\nonumber \\[1mm]  &&
+ \left\{
   t_p (t + t_p) \frac{\nu (\nu - t) }{4 (\nu^2 - \nu_0^2)}
    - t_1 t_2
\right\} \mtx0112 +
(\nu - t) t_m \left\{ \frac{1}{2} + 
      \frac{\nu t_p}{4 (\nu^2 - \nu_0^2) }
\right\} \mtx0{-1}10 +
\nonumber \\[1mm]  &&         
+ \frac{( 4 M^2 - t )(\nu - t) }{\nu^2 - \nu_0^2} t_1 t_2
\mtx2110 +
   \frac{t ( 4 M^2 - t ) ( 2 M^2 + \nu  - t ) }
   {4 M^2 (\nu^2 - \nu_0^2) } t_1 t_2
\mtx0002
\\[5mm]
 &&A_3 = (\nu - t) t_p
     \left\{ \frac{t_p ( 6 M^2 + \nu  - 3 t ) - t (\nu - t) }
     {4 (\nu^2 - \nu_0^2) }
     - t t_p \frac{( 3 M^2 - t ) \nu + ( 4 M^2 - t ) ( M^2 - t ) }
       {(\nu^2 - \nu_0^2)^2}
\right\} \mtx0002 -
\nonumber \\[1mm]  &&
   - \nu t_1 t_2 \frac{2 M^2 + \nu  - t}{\nu^2 - \nu_0^2}
\mtx0002 +
   t_p (t + t_p) \frac{M^2 (\nu - t)}{\nu^2 - \nu_0^2}
\mtx0110 +
\frac{M^2 (\nu - t) t_m t_p}{\nu^2 - \nu_0^2}
\mtx0{-1}10 +
\nonumber \\[1mm]  &&
+ \left\{
   4 M^2 t_1 t_2 \frac{4 M^2 + \nu  - t}{\nu^2 - \nu_0^2} - 
    2 M^2 (\nu - t)^2 t_p
     \frac{( 4 M^2 + \nu  - t) t_p + \nu^2 - \nu_0^2}
       {(\nu^2 - \nu_0^2)^2}
\right\} \mtx2110
\end{eqnarray}
\section{}\label{AppB}
It is convenient to use Breit frame, in which
\begin{equation}
 q = (0,0,0,\sqrt{-t}),\qquad
 P = \left( \tfrac{1}{2}\sqrt{4M^2-t},0,0,0 \right),\qquad
 K = \frac{1}{2\sqrt{4M^2-t}} \left( \nu,\sqrt{\nu^2-\nu_0^2},0,0 \right)
\end{equation}
At $\nu=\nu_0$ all three vectors have only time- and $z$- components.
The components $p''_0$ and $p''_z$ of the vector $p''$ can be expressed
via $t_1$, $t_2$ and $p''^2$ as
\begin{equation}
 p''_z = \frac{t_2-t_1}{2\sqrt{-t}},\qquad
 p''_0 = \frac{1}{2\sqrt{4M^2-t}}(2p''^2+2M^2-t_1-t_2)
\end{equation}
The following identity holds
\begin{eqnarray} \label{id}
 & &  \frac{s(t_1+t_2-t)}{(p''^2-M^2)(k''^2-m^2)}
      - \frac{s-M^2+m^2}{k''^2-m^2}
      - \frac{s+M^2-m^2}{p''^2-M^2} = \\
 & & \frac{\nu^2-\nu_0^2}{(p''^2-M^2)(k''^2-m^2)}
   \left( -\frac{1}{4}+\frac{1}{2\sqrt{4M^2-t}}
   \left[ p''_0 -\frac{\nu+4M^2-t}{\sqrt{\nu^2-\nu_0^2}}p''_x \right] \right)
 \nonumber
\end{eqnarray}
For $\nu=\nu_0$ the r.h.s. vanishes.  
Multiplying the obtained equation by arbitrary function $f(p'')$
and integrating over $d^4 p''$ we obtain the first sought relation
\begin{equation}
 \int f(p'')d^4 p'' \left. \left\{ \frac{s(t_1+t_2-t)}{(p''^2-M^2)(k''^2-m^2)}
   - \frac{s-M^2+m^2}{k''^2-m^2} - \frac{s+M^2-m^2}{p''^2-M^2} \right\}
   \right|_{\nu=\nu_0} = 0
\end{equation}
To find the relations containing the derivative of $I_4$,
we divide Eq.(\ref{id}) by $\nu^2-\nu_0^2$, multiply by arbitrary
function of the form $f(p''^2,t_1,t_2)$ and integrate over $d^4 p''$,
keeping in mind to put $\nu=\nu_0$ afterwards.
The r.h.s. will consist of three integrals, the last of which is
\begin{equation}
 \frac{1}{\sqrt{\nu^2-\nu_0^2}} \int \frac{f(p''^2,t_1,t_2)}{p''^2-M^2}
 \, \frac{p''_x d^4 p''}{s+p''^2+2K_xp''_x-2(P_0+K_0)p''_0}
\end{equation}
(the long expression in the denominator is equal to $k''^2-m^2$).
Substituting $p''_x \to -p''_x$ and averaging obtained and initial integrals,
we obtain in the limit $\nu\to \nu_0$ (which implies $K_x\to 0$)
\begin{equation}
 \left. \frac{1}{\sqrt{\nu^2-\nu_0^2}}
   \int \frac{f(p''^2,t_1,t_2)}{p''^2-M^2}
   \, \frac{p''_x d^4 p''}{k''^2-m^2} \right|_{\nu=\nu_0} = 
 - \left. \frac{1}{\sqrt{4M^2-t}} \int
   \frac{f(p''^2,t_1,t_2)}{p''^2-M^2}
   \, \frac{p''^2_x d^4 p''}{(k''^2-m^2)^2} \right|_{\nu=\nu_0} 
\end{equation}
After some algebra we obtain the second sought relation
(the electron mass $m$ was neglected in the numerators)
\begin{eqnarray}
 & & \int f(p''^2,t_1,t_2) d^4 p'' \left. \left\{ \frac{s(t_1+t_2-t)}{(p''^2-M^2)(k''^2-m^2)}
   - \frac{s-M^2+m^2}{k''^2-m^2} - \frac{s+M^2-m^2}{p''^2-M^2} \right\} \frac{t_1+t_2-t}{\nu^2-\nu_0^2}
   \right|_{\nu=\nu_0} = \\ 
 & & = \frac{1}{4\nu t} \int f(p''^2,t_1,t_2) d^4 p'' 
        \left\{ \frac{t_1+t_2-t}{\nu-t} \left[ \frac{\nu^2+t^2}{p''^2-M^2} 
              - \frac{2\nu t}{k''^2-m^2} \right] 
  - \frac{2 \nu\, t_1 t_2}{(k''^2-m^2)(p''^2-M^2)}
              + \frac{2 t\, t_1 t_2}{(k''^2-m^2)^2}
        \right\} \nonumber
\end{eqnarray}

\def\ea{{ \it et al.}}
\def\Jou#1#2#3#4#5{#1, #2 {\bf #3}, #4 (#5)}

\end{document}